\documentstyle[12pt,epsf]{article}
\begin{document}

\catcode`@=11
\long\def\@caption#1[#2]#3{\par\addcontentsline{\csname
  ext@#1\endcsname}{#1}{\protect\numberline{\csname
  the#1\endcsname}{\ignorespaces #2}}\begingroup
    \small
    \@parboxrestore
    \@makecaption{\csname fnum@#1\endcsname}{\ignorespaces #3}\par
  \endgroup}
\catcode`@=12
\newcommand{\newc}{\newcommand}
\newc{\gsim}{\lower.7ex\hbox{$\;\stackrel{\textstyle>}{\sim}\;$}}
\newc{\lsim}{\lower.7ex\hbox{$\;\stackrel{\textstyle<}{\sim}\;$}}
\newc{\gev}{\,{\rm GeV}}
\newc{\mev}{\,{\rm MeV}}
\newc{\ev}{\,{\rm eV}}
\newc{\kev}{\,{\rm keV}}
\newc{\tev}{\,{\rm TeV}}
\def\tr{\mathop{\rm tr}}
\def\Tr{\mathop{\rm Tr}}
\def\Im{\mathop{\rm Im}}
\def\Re{\mathop{\rm Re}}
\def\bR{\mathop{\bf R}}
\def\bC{\mathop{\bf C}}
\def\lie{\mathop{\hbox{\it\$}}} 
\newc{\sw}{s_W}
\newc{\cw}{c_W}
\newc{\swsq}{s^2_W}
\newc{\swsqb}{s^2_W}
\newc{\cwsq}{c^2_W}
\newc{\cwsqb}{c^2_W}
\newc{\Qeff}{Q_{\rm eff}}
\newc{\fpf}{{\bf\bar5}+{\bf5}}
\newc{\tpt}{{\bf\overline{10}}+{\bf10}}
%
%
\def\NPB#1#2#3{Nucl. Phys. {\bf B#1} (19#2) #3}
\def\PLB#1#2#3{Phys. Lett. {\bf B#1} (19#2) #3}
\def\PLBold#1#2#3{Phys. Lett. {\bf#1B} (19#2) #3}
\def\PRD#1#2#3{Phys. Rev. {\bf D#1} (19#2) #3}
\def\PRL#1#2#3{Phys. Rev. Lett. {\bf#1} (19#2) #3}
\def\PRT#1#2#3{Phys. Rep. {\bf#1} (19#2) #3}
\def\ARAA#1#2#3{Ann. Rev. Astron. Astrophys. {\bf#1} (19#2) #3}
\def\ARNP#1#2#3{Ann. Rev. Nucl. Part. Sci. {\bf#1} (19#2) #3}
\def\MPL#1#2#3{Mod. Phys. Lett. {\bf #1} (19#2) #3}
\def\ZPC#1#2#3{Zeit. f\"ur Physik {\bf C#1} (19#2) #3}
\def\APJ#1#2#3{Ap. J. {\bf #1} (19#2) #3}
\def\AP#1#2#3{{Ann. Phys. } {\bf #1} (19#2) #3}
\def\RMP#1#2#3{{Rev. Mod. Phys. } {\bf #1} (19#2) #3}
\def\CMP#1#2#3{{Comm. Math. Phys. } {\bf #1} (19#2) #3}
\relax
%
%
%
\def\beq{\begin{equation}}
\def\eeq{\end{equation}}
\def\bea{\begin{eqnarray}}
\def\eea{\end{eqnarray}}
%
%
%
\def\boxeqn#1{\vcenter{\vbox{\hrule\hbox{\vrule\kern3pt\vbox{\kern3pt
\hbox{${\displaystyle #1}$}\kern3pt}\kern3pt\vrule}\hrule}}}
%
%
\def\mbox#1#2{\vcenter{\hrule \hbox{\vrule height#2in
\kern#1in \vrule} \hrule}}
\def\half{{\textstyle{1\over2}}} 
%
%
%
%
\newc{\ie}{{\it i.e.}}          \newc{\etal}{{\it et al.}}
\newc{\eg}{{\it e.g.}}          \newc{\etc}{{\it etc.}}
\newc{\cf}{{\it c.f.}}
%
%
%
%
\def\CAG{{\cal A/\cal G}} 
\def\CA{{\cal A}} \def\CB{{\cal B}} \def\CC{{\cal C}} \def\CD{{\cal D}}
\def\CE{{\cal E}} \def\CF{{\cal F}} \def\CG{{\cal G}} \def\CH{{\cal H}}
\def\CI{{\cal I}} \def\CJ{{\cal J}} \def\CK{{\cal K}} \def\CL{{\cal L}}
\def\CM{{\cal M}} \def\CN{{\cal N}} \def\CO{{\cal O}} \def\CP{{\cal P}}
\def\CQ{{\cal Q}} \def\CR{{\cal R}} \def\CS{{\cal S}} \def\CT{{\cal T}}
\def\CU{{\cal U}} \def\CV{{\cal V}} \def\CW{{\cal W}} \def\CX{{\cal X}}
\def\CY{{\cal Y}} \def\CZ{{\cal Z}}
%
%
%
%
%
\def\grad#1{\,\nabla\!_{{#1}}\,}
\def\gradgrad#1#2{\,\nabla\!_{{#1}}\nabla\!_{{#2}}\,}
\def\partder#1#2{{\partial #1\over\partial #2}}
\def\secder#1#2#3{{\partial^2 #1\over\partial #2 \partial #3}}
%
%
%
%
%
\def\ltap{\ \raise.3ex\hbox{$<$\kern-.75em\lower1ex\hbox{$\sim$}}\ }
\def\gtap{\ \raise.3ex\hbox{$>$\kern-.75em\lower1ex\hbox{$\sim$}}\ }
\def\gl{\ \raise.5ex\hbox{$>$}\kern-.8em\lower.5ex\hbox{$<$}\ }
\def\roughly#1{\raise.3ex\hbox{$#1$\kern-.75em\lower1ex\hbox{$\sim$}}}
%
%
%
%
\def\slash#1{\rlap{$#1$}/} 
\def\dsl{\,\raise.15ex\hbox{/}\mkern-13.5mu D} 
\def\delsl{\raise.15ex\hbox{/}\kern-.57em\partial}
\def\Ksl{\hbox{/\kern-.6000em\rm K}}
\def\Asl{\hbox{/\kern-.6500em \rm A}}
\def\Dsl{\hbox{/\kern-.6000em\rm D}} 
\def\Qsl{\hbox{/\kern-.6000em\rm Q}}
\def\gradsl{\hbox{/\kern-.6500em$\nabla$}}
%
%
\let\al=\alpha
\let\be=\beta
\let\ga=\gamma
\let\Ga=\Gamma
\let\de=\delta
\let\De=\Delta
\let\ep=\varepsilon
\let\ze=\zeta
\let\ka=\kappa
\let\la=\lambda
\let\La=\Lambda
\let\del=\nabla
\let\si=\sigma
\let\Si=\Sigma
\let\th=\theta
\let\Up=\Upsilon
\let\om=\omega
\let\Om=\Omega
\def\ph{\varphi}
%
%
%
\newdimen\pmboffset
\pmboffset 0.022em
\def\oldpmb#1{\setbox0=\hbox{#1}%
 \copy0\kern-\wd0
 \kern\pmboffset\raise 1.732\pmboffset\copy0\kern-\wd0
 \kern\pmboffset\box0}
\def\pmb#1{\mathchoice{\oldpmb{$\displaystyle#1$}}{\oldpmb{$\textstyle#1$}}
	{\oldpmb{$\scriptstyle#1$}}{\oldpmb{$\scriptscriptstyle#1$}}}
%
%
%
%
%
\def\bar#1{\overline{#1}}
\def\vev#1{\left\langle #1 \right\rangle}
\def\bra#1{\left\langle #1\right|}
\def\ket#1{\left| #1\right\rangle}
\def\abs#1{\left| #1\right|}
\def\vector#1{{\vec{#1}}}
\def\inv{^{\raise.15ex\hbox{${\scriptscriptstyle -}$}\kern-.05em 1}}
\def\pr#1{#1^\prime}  
\def\lbar{{\lower.35ex\hbox{$\mathchar'26$}\mkern-10mu\lambda}} 
\def\e#1{{\rm e}^{^{\textstyle#1}}}
\def\ee#1{\times 10^{#1} }
\def\om#1#2{\omega^{#1}{}_{#2}}
\def\imp{~\Rightarrow}
\def\coker{\mathop{\rm coker}}
\let\p=\partial
\let\<=\langle
\let\>=\rangle
\let\ad=\dagger
\let\txt=\textstyle
\let\h=\hbox
\let\+=\uparrow
\let\-=\downarrow
\def\dot{\!\cdot\!}
\def\vfilll{\vskip 0pt plus 1filll}
%

\begin{titlepage}
{\hbox to\hsize{{\large hep-ph/9606384} \hfill{\large IASSNS-HEP-96/63}}}\par
{\hbox to\hsize{{\large June 1996} \hfill{\large BA-96-23}}}\par

\vskip 2cm
\begin{center}
{\Large\bf Gauged SO(3) Family Symmetry and}\\ \vskip .2cm
{\Large\bf Squark Mass Degeneracy\footnote{Research
supported in part by DOE grants DE-FG02-90ER40542 and DE-FG02-91ER406267.
Email: {\tt babu@sns.ias.edu, smbarr@bartol.udel.edu}}}
\vskip 1cm
{\large
K.S.~Babu$^*$
and S.M. Barr$^\dagger$\\}
\vskip 0.5cm
{\large\sl $^*$School of Natural Sciences\\
Institute for Advanced Study\\
Princeton, NJ~08540\\} \vskip .3cm
{\large\sl $^\dagger$Bartol Research Institute\\
University of Delaware\\
Newark, DE 19716\\}
\end{center}
\vskip .5cm

\begin{abstract}

It is shown that a gauged SO(3) family symmetry can suppress
flavor-changing processes from squark-mass non-degeneracy to 
an acceptable level.  The potentially dangerous SO(3) D-terms 
can be rendered harmless if the CP-violating phases appearing 
in the SO(3)-breaking sector are small, which can naturally be 
the case if CP is a spontaneously broken symmetry.  This approach 
has certain advantages over models based on global or
non-abelian discrete symmetries, and dovetails with some
recent proposals for explaining the pattern of quark and lepton
masses. Moreover this approach relates the near-degeneracy of the
squark masses to an approximate CP invariance which can also explain 
the smallness of the electron and neutron electric dipole moments.

\end{abstract}

\end{titlepage}
\setcounter{footnote}{0}
\setcounter{page}{1}
\setcounter{section}{0}
\setcounter{subsection}{0}
\setcounter{subsubsection}{0}

\baselineskip= .25in

\section{Introduction}

Non-degeneracy of squark masses can lead to ``flavor-changing
neutral currents" (FCNC) that are unacceptably large \cite{dimgeo}.  
In particular, the limits on the real and imaginary parts
of the mass splitting between the $K_L$ and $K_S$ suggest
a near degeneracy between the $\tilde{d}$ and $\tilde{s}$ 
squarks \cite{dono,maseiro}.  Such a near
degeneracy is hard to understand in the absence of some symmetry
principle. The problem is made even more obscure by the fact
that the quarks, far from exhibiting such a degeneracy, have
a dramatic hierarchy of masses.

A symmetry rendering $\tilde{s}$ and $\tilde{d}$ nearly degenerate must
be non-abelian.\footnote{Degeneracy of squark masses could also arise if
supersymmetry breaking is communicated to the squark sector via gauge
interactions.  For models based on this idea, see Ref. \cite{dine}.
In this paper we shall be concerned with supersymmetry breaking via
supergravity, in which case the squarks are not degenrate in general.   
For attempts to solve the squark--mediated FCNC
problem using Abelian global symmetries, see Ref. \cite{nir}.}   
It has been proposed in the literature 
\cite{pouliot,pomarol,hall}
that this role can be played by a family SU(2), or a discrete 
subgroup of it, under which the lightest two families form
a doublet and the third generation forms a singlet. This is a
structure consonant with both the presumed approximate 
degeneracy of $\tilde{s}$ and $\tilde{d}$ and the relatively large masses 
of $t$, $b$, and $\tau$. 

An SU(2) under which the families form a ${\bf 2 + 1}$ cannot
be gauged, as then the D-terms of SU(2) would directly
contribute splittings to the squark masses comparable to
the squark masses themselves. On the other hand, various
arguments based on quantum gravity
\cite{banks} seem to disfavor a global continuous symmetry below the
Planck or string scale. This has led to the consideration
of non-abelian discrete symmetry groups \cite{frampton,pomarol}.   
However, since non-abelian discrete symmetries have up to now played little
if any role in explaining particle phenomena, one may wonder if
this is a completely satisfactory approach.

In this letter we shall discuss the possibility that the
three families form a triplet of a gauged SO(3). Because
the representations of SO(3) are real it turns
out that the troublesome
D-terms vanish for real vacuum expectation values. If CP
is spontaneously broken then the SO(3)-breaking VEVs can,
indeed, naturally be approximately real. The squark-mass
splittings then would be controlled by a naturally small CP-violating phase.
This is very appealing since there are independent grounds for
suspecting that certain CP-violating phases must be small
in supersymmetric theories. In particular, the one-loop
contributions to quark and electron electric dipole moments are 
small \cite{ellis}.   Moreover, the severest constraint on the non-degeneracy
of $\tilde{d}$ and $\tilde{s}$ --- namely from the $\epsilon_K$
parameter --- is greatly alleviated if the relevant CP-violating
phase is small \cite{dono,maseiro,pouliot,pomarol,hall,frampton}.  

In the kinds of models we shall discuss, the large third-generation
quark and lepton masses arise from a VEV which breaks SO(3)
down to an approximate SO(2). As will be seen, in spite of
the fact that SO(2) is abelian, the splitting of $\tilde{s}$ 
and $\tilde{d}$ is sufficiently controlled. Indeed, since the families
form a ${\bf 2 + 1}$ (of course, reducible) under this residual 
SO(2), much of the
discussion of squark mass splittings carries over with little
change from the
cases of SU(2) and its subgroups appearing in the literature.

Some recently proposed models \cite{barr} have shown that quark and lepton
masses can be accounted for without using symmetries that
distinguish among the generations --- all generations 
are treated
on an equal footing \cite{mohap}. In these models certain Yukawa couplings are
vectors in ``family space", that is they carry one family index.
These models can be transposed very naturally into the context of
a gauged family SO(3). In that transposition the Yukawa couplings
that were vectors in family space are replaced by Higgs triplets
of the SO(3). We will present an illustrative model that has a
close affinity with that of Ref. \cite{barr}, and which we believe shows
certain advantages of SO(3) over other groups that have been discussed.
In the model discussed in Section 4.2, most of the observed features
of the quark and lepton masses are explained by various aspects of
SO(10) gauge symmetry, while the squark mass degenracy is understood
with a (commuting) SO(3) family gauge symmetry.  

\section{The FCNC Problem and Squark-mass Degeneracy}

Introducing low-energy supersymmetry produces several
problems \cite{dimgeo,dono,maseiro,pouliot,ellis,rata} 
connected with flavor violation and CP violation
that are absent in the Standard Model. For purposes of
discussion it is convenient to distinguish four of them,
even though in particular models they are closely related.

\begin{itemize}

\item (1) If the $\tilde{s}$ and $\tilde{d}$ masses are non-degenerate then mixing
of $s$ and $d$ will lead, in general, to an excessively large
$K_L$-$K_S$ mass splitting, $\Delta m_K$. For maximal squark 
non-degeneracy Re$M_{12}$ comes out about $10^4$ times too 
large \cite{dono,maseiro}.  There are somewhat less dangerous contributions
(by about an order of magnitude) also from
mixing of $\tilde{b}$ with $\tilde{d}$.

\item (2) The same effect, if CP-violating phases are present, leads
to an excessively large Im$M_{12}$, that is to a too large 
$\epsilon_K$. If the relevant phases are of order unity,
then $\epsilon_K$ comes out about $10^7$ times too big \cite{dono,maseiro}.
(The phases that are of most concern are those in the squark-mass
matrix, although it should be kept in mind that CP--phases in the
quark-mass matrix can show up in the gluino box diagram
when one rotates the phase of the $s$ quark to make the
$K^0 \rightarrow \pi \pi (I = 0)$ amplitude real.)

\item (3) Even with ``universal" soft SUSY-breaking terms ({\it i.e.}
degenerate squark masses and proportionality of $A$ terms) the
electric dipole moment of the neutron will receive contributions
at one (gluino) loop if the $A$ parameter or the $\mu$ parameter
are complex. If these phases are of order unity, $d_n$ comes
out typically about $10^2$ times too big \cite{ellis,maseiro}.  

\item (4) Finally, even if $A$ and $\mu$ are real, non-proportionality
of the $A$-terms of the squarks to the quark Yukawa couplings
gives a one-gluino-loop contribution to $d_n$ if there are
phases in the quark mass matrix. If these phases are of
order unity (as they need to be
for the Kobayashi-Maskawa mechanism to work) 
typically $d_n$ is about $10^2$ too big \cite{rata}.  
\end{itemize}

It should be noted that these problems have somewhat different
origins. Problems (1) and (2) arise from non-degeneracy of squark
masses. Problems (2) and (3) arise from CP-violating phases
in the soft terms that break supersymmetry.  
Problem (4) comes from non-proportionality of the $A$ terms.

It is with problems (1) and (2) that we are primarily concerned
in this paper. They are quantitatively the most severe and 
would seem to suggest a high degree of degeneracy of the squarks
(especially $\tilde{d}$ and $\tilde{s}$) no matter how
CP is broken in the theory. (Indeed, problem (1) is a CP-conserving effect.)
Our purpose in this paper is to propose an attractive explanation
for this near-degeneracy. This explanation entails that certain
CP-violating phases be small, which in turn suggests that CP is
spontaneously broken. This fits in well with the requirements
for solving problems (2) and (3), namely that the phases
in the soft terms be small. However, neither squark degeneracy
nor small phases in the soft terms will solve problem (4).
The solution of problem (4) requires further conditions on the theory,
in particular on the sector that generates quark and lepton masses.
This issue will be discussed in Section 6. For most of this
paper, when we discuss the FCNC problem we will mean problems
(1) and (2). 

It should be noted that even if the squarks are non--degenerate at a
high scale $M_{\rm GUT}$ (or $M_{\rm Pl}$), some amount of degeneracy could
arise in the process of renormalization group running from $M_{\rm
GUT}$ to the weak scale.  This is a consequence of the gaugino loop
corrections to the squark masses, which are flavor--blind.  Similarly,
the $A$ terms could become proportional to the Yukawa couplings to
some extent as a result of the RGE flow, even though they are
non--proportional at $M_{\rm GUT}$.  The squark masses and the
$A$ parameters at the weak scale (for the lighter two families) are given by
\begin{eqnarray}
m_{Q_i}^2(m_Z) &\simeq & m_{Q_i}^2(M_{\rm GUT}) + 7.2 M_{1/2}^2 \nonumber
\\
A_i(m_Z) & \simeq & A_i(M_{\rm GUT}) -4.1 M_{1/2}~.
\end{eqnarray}
A 10\% degeneracy in the squark mass--squared is not unreasonable to
expect from the RGE flow (e.g: if $M_{Q_i}(M_{\rm GUT}) \sim
M_{1/2}$), but this by itself is not sufficient to explain the
constraints from the Kaon system.  Similarly a 10\% proportionality of
the $A$ terms and the Yukawa couplings is quite plausible (e.g: if
$A_i(M_{\rm GUT}) \lsim M_{1/2}$).  Since problem (4) noted above
requires this proportionality to hold only to about 1\%, it appears to
us that (4) is a much less severe problem.  

The FCNC problem puts the most severe constraints on the degeneracy
of the $\tilde{s}$ and $\tilde{d}$. Let us define $m_Q$ and $m_D$ to be the
$LL$ and $RR$ mass matrices of the down squarks, respectively.
And let $V^Q$ and $V^D$ be the unitary transformations that take
the squark mass matrices from the basis where the quark mass
matrices are real and diagonal and where the gluino couplings are
flavor-conserving to the basis where $m_Q$ and $m_D$ are real
and diagonal. Then the constraint from the $K_L$-$K_S$ mass
splitting is (for $m_Q \simeq m_{\tilde{g}}$) \cite{maseiro}
\begin{equation}
\left( \frac{{\rm TeV}}{m_Q} \right)^2 \left|
\frac{V^Q_{11} \delta m^2_Q V^Q_{21}}{m^2_Q}
\frac{V^D_{11} \delta m^2_D V^D_{21}}{m^2_D} \right| < 3.1 \times
10^{-3}.
\end{equation}
\noindent
The constraint from $\epsilon_K$ is that
\begin{equation}
\left( \frac{{\rm TeV}}{m_Q} \right)^2 \left|
\frac{V^Q_{11} \delta m^2_Q V^Q_{21}}{m^2_Q}
\frac{V^D_{11} \delta m^2_D V^D_{21}}{m^2_D} \right| \sin \phi < 1.9 \times
10^{-7},
\end{equation}
\noindent
where $\phi \equiv \arg(V^Q_{11} V^{Q*}_{21} V^D_{11} V^{D*}_{21})$.

From mixing with $\tilde{b}$ one has the condition (again from
$\epsilon_K$) that
\begin{equation}
\left( \frac{1 \; {\rm TeV}}{m_{\tilde{b}}} \right)^2
\left| V^Q_{13}V^Q_{23}V^D_{13}V^D_{23} \right| \sin \phi'
< 5 \times 10^{-8}.
\end{equation}

\noindent
If $\tilde{d}$ and $\tilde{s}$ had maximal non-degeneracy, that is
if $\delta m_{Q,D}^2/m_{Q,D}^2 \simeq 1$, and the phase
$\phi$ were of order unity, then Eq. (3) would 
require that $m_Q \stackrel{_>}{_\sim} 200~ {\rm TeV}$.

In the models under discussion in this paper (and also those in
Ref. \cite{pomarol}) one has that
\begin{equation}
m_Q^2, m_D^2  = \left( 
\begin{array}{ccc}
1 & \lambda^3 & \lambda^3 \\
\lambda^3 & 1 + \lambda^2 & \lambda^2 \\
\lambda^3 & \lambda^2 & O(1) 
\end{array}
\right) m^2,
\end{equation}

\noindent
where $\lambda^n$ means O($\lambda^n$), and where $\lambda = {\rm
sin}\theta_C \simeq .22$
is the Wolfenstein parameter. That means that in Eq. (2),
$\delta m_{Q,D}^2/m_{Q,D}^2 \sim \lambda^2$ and 
$V^{Q,D}_{21} \sim \lambda$, and thus the condition there
is easily satisfied for squark masses of order 100 GeV.
The condition from the $\epsilon_K$ parameter, however,
gives (Eq. (3)) that $m_Q > 10^3 \lambda^3 \sqrt{ \sin \phi}~
{\rm TeV} \simeq 24~ {\rm TeV}~ \sqrt{ \sin \phi}$, which
implies that $\phi$ is of order $10^{-3}$ or $10^{-4}$.
And in Eq. (4), $V^{Q,D}_{13} \sim \lambda^3$, $V^{Q,D}_{23}
\sim \lambda^2$, so that $m_Q > 4.5 \times 10^3 \lambda^5
\sqrt{ \sin \phi'}~ {\rm TeV} \sim 2.3 \sqrt{ \sin \phi'}~ 
{\rm TeV}$. Thus $\phi'$ must also be small.
(As noted parenthetically where ``problem (2)" was defined
above, phases in the quark-mass matrix also contribute
to $\epsilon_K$ indirectly through the gluino box
diagram. In particular, the most dangerous contribution comes
from a phase in the (12) or (21) elements of the quark-mass
matrices. These also give the most dangerous contribution
to problem (4). Thus whatever solves problem (4) tends to
eliminate or lessen this problem as well. See the discussion in
Sec. 6.)

To see how such a form as Eq. (5) may be achieved using
non-abelian family symmetry, it is helpful to review,
at least in outline, a model proposed in Ref. \cite{pomarol}. 

Consider a model with low-energy supersymmetry and the 
family group $G_H$, which is either $SU(2) \times U(1) \times
Z_2$ or some discrete subgroup thereof. The families are
arranged in ${\bf 2 + 1}$ representations of the non-abelian
group: $(Q + Q_3)$, $(d^c + d^c_3)$, and $(u^c + u^c_3)$, where
the representations without the subscript `3' are understood to be 
doublets under the family group, and to contain the first two generations.
Breaking the non-abelian 
family group is a doublet, $\Phi$, which has VEV given by
$\langle \Phi \rangle = \Lambda (0, \epsilon)^T$. 
$\Lambda$ is a scale characterizing
new physics that is represented by higher-dimensional operators
involving $\Phi$ appearing in an effective lagrangian. 
The doublets, $Q$, $d^c$, and $u^c$, all have charge $1$ under
the U(1), while the doublet $\Phi$ has charge $-1$. The Higgs
field $\Phi$ is odd under the $Z_2$ as is a singlet Higgs field $\chi$.

The third generation fermions are heavy because their masses
arise from dimension-four terms: 
\begin{equation}
{\cal L}_Y^0 = h_d d^c_3 H_d Q_3 + h_u u^c_3 H_u Q_3.
\end{equation}

\noindent
The next layer of the ``onion structure" of the mass matrices
is provided by the higher-dimension terms 
\begin{equation}
{\cal L}_Y = h_d (d^{cT}, d^c_3) \left( 
\begin{array}{cc}
h^d_{\Phi \Phi} \Phi \Phi^T/\Lambda^2 & h^d_{\Phi \chi} 
\Phi \chi/\Lambda^2 \\ h^d_{\chi \Phi} \Phi^T \chi/\Lambda^2 & 1 \\
\end{array} \right) \left( \begin{array}{c} Q \\ Q_3 \end{array}
\right) H_d + \{ d \rightarrow u \} + h.c..
\end{equation}

\noindent
The singlet field $\chi$ has VEV 
$\langle \chi \rangle = \Lambda \epsilon_{\chi}$,
where $\epsilon_{\chi}$ is of the same order as $\epsilon$.
This gives the following structure to the down-quark mass
matrix;

\begin{equation}
M_d = \left( 
\begin{array}{ccc}
0 & 0 & 0 \\
0 & h^d_{\Phi \Phi} \epsilon^2 & h^d_{\Phi \chi} \epsilon \epsilon_{\chi} \\
0 & h^d_{\chi \Phi} \epsilon \epsilon_{\chi} & 1 
\end{array} \right) h_d \langle H_d \rangle,
\end{equation}

\noindent
with a similar form for the up-quark and charged lepton
mass matrices. From the fact that $V_{cb}$ and
$m_s/m_b$ are roughly the same and of order $\lambda^2$ one sees that
$\epsilon \simeq \epsilon_{\chi} \sim \lambda$.  However, this fact is not
explained in this framework, since the fields $\chi$ and $\Phi$
have different symmetry properties, and therefore there is
no obvious reason why $\epsilon$ and $\epsilon_{\chi}$ should be 
comparable. 
The SO(3) models, as we
shall see, can do better in this regard.
There are several ways to generate the first generation masses 
and mixings,$^{}$
but these are not of immediate concern to us.

The leading contribution to the squark-masses come from
effective operators \cite{pomarol,grisaru} of the form
\begin{equation}
{\cal L}_{{\rm soft}} = \int d^4 \theta Q^{\dag} Q \overline{\eta} \eta,
\end{equation}

\noindent
and others with $Q$ replaced by $Q_3$, $d^c$, $d^c_3$ and so on.
$\eta$ is a dimensionless ``spurion" superfield which represents
the effect of supersymmetry breaking. $\langle \eta \rangle 
= m_{SUSY} \theta^2$. 
The family symmetry, SU(2), or a discrete non-abelian subgroup of it,
ensures that there is a degeneracy between the sparticle masses
of the first two generations.
  
There are a number of contributions to the 
non-degeneracy of $\tilde{s}$ and $\tilde{d}$. 
One is from higher-dimension operators
of the form \cite{pomarol,grisaru}
\begin{equation}
{\cal L'}_{\rm soft} = \int d^4 \theta Q^{\dag} \Phi \Phi^{\dag}
Q \overline{\eta} \eta/\Lambda^2,
\end{equation}

\noindent
From the form of the VEV of $\Phi$, and the structure of the mass
matrix $M_d$, it is readily apparent that the effect of such a term
is to generate a contribution 
to the mass--squared for $\tilde{s}$ that is of order
$\lambda^2 m_{SUSY}^2$. In other words, the fractional splitting
between the $\tilde{d}$ and $\tilde{s}$ masses is of order $10^{-2}$. 
This is true both of the splitting in the left-squark
sector (given by the expression in Eq. (10)) and of the splitting in the
right-squark sector (given by the analogous expression with $Q$
replaced by $d^c$). 
There are other contributions to the squark mass splittings, but they are
no larger in order of magnitude than this.

\section{SO(3) Breaking and the Effect of the D-Terms}

Consider a gauged SO(3) symmetry broken spontaneously
by a set of M vectors: $\vec{A}_i$, $i = 1, ..., M$.
Assume that the left and right-handed (s)quarks and (s)leptons are
all in vector representations of SO(3), which we will generically
denote by $\vec{\phi}_n$. 
The SO(3) D term in the potential then has the form
\begin{equation}
V_D(\vec{A}_i) = \sum_{a= 1}^3 (D^a)^2 = \left| \sum_n \vec{\phi}_n^* 
\times \vec{\phi}_n
+  \sum_i \vec{A}^*_i \times \vec{A}_i \right|^2.
\end{equation}

\noindent
Note that the generators of SO(3) in the vector representation
can be written using the familiar cross product. It is clear that if the
VEVs that spontaneously break SO(3) contribute non-vanishingly
to the SO(3) D-term, the cross terms in Eq. (11) will lead to 
intergenerational splittings of the squark masses. It is this 
that has made gauged continuous family symmetries
disfavored as a solution to the squark-mass non-degeneracy problem.
However, it is obvious that if CP were conserved, the VEVs
of the vectors $\vec{A}_i$ would be real and would contribute nothing
to the dangerous D-terms, since 
$\vec{A}^*_i \times \vec{A}_i = 2 i \vec{A}_{iR} \times \vec{A}_{iI}$.

Since the SO(3) D-terms break supersymmetry, they must be
proportional to $m_{SUSY}$. Thus one expects that the splittings
among the mass--squared of the squarks due to these D-terms will be of
order $\theta_{CP} m_{SUSY}^2$, where $\theta_{CP}$ 
represents the magnitude of the CP-violating
phases appearing in the potential of the vectors $\vec{A}_i$.
Since the overall magnitude of the squark masses is itself
of order $m_{SUSY}$, the D-term contributions to the fractional splittings 
$\delta m_Q^2/m_Q^2$ and $\delta m_D^2/m_D^2$ appearing
in Eqs. (2) and (3) are of order $\theta_{CP}$. In a theory in which CP
is spontaneously broken, these phases can be naturally small.

To be more explicit, let us assume that there are three vectors,
$\vec{A}_i$, $i = 1,2,3$, which break SO(3) spontaneously at a
scale $\Lambda$ which is near the Planck scale. Since we will
assume that the superpotential for these fields has CP-violating
phases which are very small, we can expand the VEVs of the
vectors $\vec{A}_i$ in 
powers of this CP violation. In particular, we can write
\begin{equation}
\langle \vec{A}_i \rangle = \vec{A}^{(0)}_i + i \vec{b}_i + O(\theta_{CP}^2),
\end{equation}
\noindent
where $\vec{A}^{(0)}_i$ is the VEV in the limit that the CP-violating
phases are set to zero. The imaginary parts of the VEVs,
$\vec{b}_i$, are thus of order $\theta_{CP}$. 

Clearly, if we neglect the soft, supersymmetry-breaking terms that
are of order the weak scale, the VEVs of the $\vec{A}_i$ must make
what remains of the potential, namely the F-terms and D-terms, vanish ---
otherwise supersymmetry would break at the scale $\Lambda \sim M_{Pl}$.
The F terms can only depend on the $\vec{A}_i$ through the six
SO(3)-invariant complex quantities $\vec{A}_i \cdot \vec{A}_j$.
Thus their vanishing gives 12 real conditions to be satisfied. 
(Another way to argue this, which applies to any number, M, of vectors,
is that the $\vec{A}_i$ contain $6M$ real quantities, but the
superpotential, being holomorphic, is invariant under
the complex extension of SO(3), and so 6 of the parameters are
left undetermined by the F terms. Thus $6(M-1)$, in our case 
12, conditions are imposed by the F terms.) 
These can be written to leading order in CP violation
\begin{equation}
\begin{array}{ccl}
\vec{A}^{(0)}_i \cdot \vec{b}_i & = & \left| \vec{A}_i^{(0)} \right|^2 
\theta_i \\
 \vec{A}^{(0)}_i \cdot \vec{b}_j + \vec{A}^{(0)}_j \cdot \vec{b}_i
& = & \left| \vec{A}^{(0)}_i \cdot \vec{A}^{(0)}_j \right| \theta_{ij}, \;\;
i \neq j,
\end{array}
\end{equation}
\noindent
where $\theta_i$ and $\theta_{ij}$ are combinations of CP-violating
phases appearing in the superpotential, and are of order
$\theta_{CP}$. 

The D-terms depend
only on the real three-vector $\vec{D}$, 
and so their vanishing gives 3 more real 
conditions, which can be written to leading order in CP-violation as
\begin{equation}
\sum_i \vec{A}^{(0)}_i \times \vec{b}_i = 0.
\end{equation}
\noindent
But the three vectors $\vec{A}_i$ have 18 real
degrees of freedom, of which three are gauge degrees of freedom.
Thus there are 15 conditions to be satisfied by 15 parameters, and
there is in general a non-trivial solution to Eqs. (13) - (14).
This solution can be expressed schematically as $b \sim \theta_{CP} A^{(0)}$.
The D-term in the potential can be written in the same schematic
way as $V_D \sim \left| g A^{(0)} (b - \theta_{CP} A^{(0)}) + g \phi_n^*
\tau \phi_n \right|^2$. $g$ is the SO(3) gauge coupling constant,
and $\tau$ is an SO(3) generator. 
When the soft supersymmetry-breaking terms are
added, one has effectively a potential for $\vec{b}_i$ that can
be written as $V(b) \sim g^2 (A^{(0)})^2 (b -  \theta_{CP} A^{(0)})^2 +
m_0^2 \left| b \right|^2$.
This gives $g^2 (A^{(0)})^2 (b - \theta_{CP} A^{(0)}) \sim m_0^2 b 
\sim \theta_{CP} m_0^2 A^{(0)}$. Therefore the contribution to the
squark and slepton mass splittings are of order
$g^2 A^{(0)}(b - \theta_{CP} A^{(0)}) (\phi_n^* \tau \phi_n) \sim 
\theta_{CP} m_0^2  (\phi_n^* \tau \phi_n)$. This verifies our
earlier claim that the fractional squark-mass splittings due
to the SO(3) D-terms are simply of order the CP-violating phases that
appear in the superpotential of the vectors that break SO(3).
Observe that the SO(3) gauge coupling drops out.
 
A more careful treatment can be done that gives the order in
$\lambda$ of the contributions to the different squark-mass-matrix
elements coming from these SO(3) D-terms. Suppose that 
$\left| \vec{A}_1 \right|^2 \sim 1$, 
$\left| \vec{A}_2 \right|^2 \sim \lambda^2$, 
and $\left| \vec{A}_3 \right|^2 \sim \lambda^3$, 
and that $\vec{A}_1$ lies approximately in the 
`3' direction in family space, $\vec{A}_2$ lies
approximately in the `2-3 plane', and $\vec{A}_3$
has components in all three directions, with the direction
cosines $\hat{A}_i \cdot \hat{A}_j$ all being of order
unity. (These conditions apply to the types of models 
described in the next section.) Then it can be shown
that the SO(3) D-term contributions to the squark masses
are of order $\theta_{CP} m_{SUSY}^2 (O(\lambda^2),
O(\lambda^3), O(\lambda)) \cdot \vec{Q}^{\dag} \times \vec{Q}$.
In particular, writing $\vec{Q} = (\tilde{d}_L, \tilde{s}_L,
\tilde{b}_L)$,
\begin{equation}
\vec{Q}^{\dag} m_Q^2 \vec{Q} =
(\tilde{d}_L, \tilde{s}_L, \tilde{b}_L)^*
\left( \begin{array}{ccc}
1 & \lambda^3 + i \theta_{CP} \lambda & 
\lambda^3 + i \theta_{CP} \lambda^3 \\
\lambda^3 - i \theta_{CP} \lambda 
& 1 + \lambda^2 & \lambda^2 + i \theta_{CP} \lambda^2 \\
\lambda^3 - i \theta_{CP} \lambda^3 & \lambda^2 - i \theta_{CP} \lambda^2
& O(1)
\end{array}
\right) \left( \begin{array}{c}
\tilde{d}_L \\ \tilde{s}_L \\ \tilde{b}_L
\end{array} \right),
\end{equation}

\noindent
where, as before, $\lambda^n$ means $O(\lambda^n)$.
The real parts come from other sources, to be discussed later.
The imaginary (CP-violating) parts come from the SO(3) D-terms.
There are similar matrices for $\vec{D}^c = (\tilde{d}^c_L,
\tilde{s}^c_L, \tilde{b}^c_L)$.

Since we know that to satisfy the condition on the real part
of $M_{12}$ in the Kaon system (Eq. (2)) the $\tilde{d}^* \tilde{s}$
and $\tilde{s}^* \tilde{d}$ elements of $m_{Q,D}^2$ must be 
$O(\lambda^3)$, one sees from Eq. (15) that $\theta_{CP}$ must
also be no larger than $O(\lambda^2)$. Moreover, from Eq. (15)
it is easy to see
that the phase $\phi$ in Eq. (3) is $\arg\{[O(\lambda^3) 
+ i O(\theta_{CP} \lambda)]^2\} \sim \theta_{CP}/\lambda^2$.
Thus, to satisfy the condition on $\epsilon_K$ one needs
$\theta_{CP} \sim 10^{-4}$.

It is noteworthy that CP-violating effects from squark
mixing involving the third generation is more highly suppressed.  The
constraint arising from the (CP--conserving) $B_d-\overline{B}_d$
system is
\begin{equation}
\left( \frac{{\rm TeV}}{m_Q} \right)^2 \left|
\frac{V^Q_{11} \delta m^2_Q V^Q_{31}}{m^2_Q}
\frac{V^D_{11} \delta m^2_D V^D_{31}}{m^2_D} \right| < 1.3 \times
10^{-3}.
\end{equation}

\noindent From the form of Eq. (15), noting that $\delta m^2_Q/m^2_Q
\sim {\cal O}(1)$ for the $\tilde{b}-\tilde{d}$ system, we see that
the quantity on the left--hand side of Eq. (16) is approximately
$\lambda^6 \sim 1 \times 10^{-4}$.  This clearly satisfies the constraint.  As
for the CP--violating effects, note that they arise with additional
factors of $\theta_{CP}$ and are thus extremely small.
 
\section{Types of Models}

We may distinguish two approaches to constructing realistic
models using an SO(3) family gauge symmetry. What we will call
Approach I is similar in effect to models with abelian continuous
or discrete family symmetries that distinguish among the generations.  
This approach can lead to ``texture'' models of fermion masses.  
The main difference here is that in our case we are using gauged SO(3)
symmetry to arrive at such forms.  
Approach II is similar in effect to models based on ``factorization"
of Yukawa terms \cite{barr1,mohap,barr}.

\subsection{\bf Approach I: Texture models}

Let there be a set of vectors which break SO(3) spontaneously.
We will call them $\vec{A}$, $\vec{B}$, $\vec{C}$, etc.
(instead of $\vec{A}_i$ as heretofore). One can imagine
that the superpotential of these vectors is such as to
make the VEVs of any pair of them either orthogonal or
parallel.\footnote{
Fairly simple superpotentials lead to this result.
For example, the superpotential $W = \sum_{i=1}^3 \vec{A}_i\cdot \vec{A}_i +
\sum_{i,j=1}^3(\vec{A}_i\cdot \vec{A}_j)^2$ has a SUSY preserving
minimum where $\vec{A}_3 \sim (0,0,1),~\vec{A}_2 \sim
(0,1,0),~\vec{A}_1 \sim (1,0,0)$.}
Each vector can be taken to define a direction in
``family space" corresponding to a particular generation
(before diagonalizing the fermion mass matrices). For example,
$\vec{A}$ might be the longest vector and be involved in giving
mass to the heavy third generation through effective terms like
$\vec{A} \cdot \vec{D}^c \vec{A} \cdot \vec{Q} H$, etc.
Thus $\vec{A}$ would define the `3' direction in family space,
and this term would be $A^2 d^c_3 Q_3 H$. Since the VEVs of $\vec{B}$
and $\vec{C}$ are orthogonal to that of $\vec{A}$ and to each other,
they can be taken to define the `2' and `1' directions in
family space. Thus a (21) element would come from $\vec{B} \cdot
\vec{D}^c \vec{C} \cdot \vec{Q} H = BC(d^c_2 Q_1)H$.

A set of discrete symmetries that distinguish
among the vectors $\vec{A}$, $\vec{B}$, $\vec{C}$, etc. would
effectively be equivalent to a set of discrete symmetries that
distinguish family generation number. Consider, for
example, a set of $Z_2$ symmetries, $K_A \times K_B \times
K_C$, where $K_A: \vec{A} \longrightarrow - \vec{A}, \;
\vec{B} \longrightarrow + \vec{B}, \; \vec{C} \longrightarrow
+ \vec{C}$, and similarly for $K_B$ and $K_C$. If one had
a scalar field that had parity $(-,-,+)$ under
$K_A \times K_B \times K_C$ then it could couple 
to $\vec{B} \cdot \vec{D}^c \vec{A} \cdot \vec{Q}$
and $\vec{A} \cdot \vec{D}^c \vec{B} \cdot \vec{Q}$ and therefore
in the (23) and (32) elements of mass matrices but not in others. 
If this scalar is a GUT non--singlet (such as a ${\bf 45}$
of SO(10)), it will not upset the desired VEV pattern of the three
SO(3) vectors.  
In such a way ``texture" models could be constructed
of the quark and lepton mass matrices. So, although one does
not actually have an abelian family symmetry $Z_2
\times Z_2 \times Z_2$ one achieves the same result.
More complicated discrete family symmetries can be
simulated in the same way.

It should also be noted that the kind of discrete symmetry
suggested above could explain why the vectors $\vec{A}$,
$\vec{B}$, $\vec{C}$, etc., have VEVs that are orthogonal
or parallel. For example, a term in the superpotential
of the form $M \vec{A} \cdot \vec{B}$ would be forbidden,
and $(\vec{A} \cdot \vec{B})^2/M$ allowed. If terms
of yet higher dimension can be neglected then $\theta_{AB}
= n \pi/2$. On the other hand, if both these terms were present
$\theta_{AB}$ could take non-trivial values.

As an example of an SO(10) model in this approach, consider the
couplings 
\begin{eqnarray}
W = (\vec{\bf 16}\cdot \vec{A})\overline{\bf 16}_A + (\vec{\bf
16}\cdot \vec{B})\overline{\bf 16}_B + (\vec{\bf 16}\cdot
\vec{C})\overline {\bf 16}_C + \sum_{\alpha=A,B,C}\overline{\bf 16}_\alpha{\bf 16}_\alpha +
\sum_{\alpha=A,B,C}{\bf 16}_\alpha {\bf 16}_\alpha {\bf 10} \nonumber
\end{eqnarray}
where $\overline{\bf 16}_A$ carries $K_A$ charge etc.  For generation
mixing, we utilize the couplings ${\bf \overline{16}}_A {\bf 16}_B
{\bf 45}_{AB}$ where ${\bf 45}_{AB}$ has both $K_A$ and $K_B$ charges, etc.  
When the ${\bf \overline{16}}_{A,B,C}+ {\bf
16}_{A,B,C}$ are integrated out, light fermion mass matrices with
specific textures will emerge.

\subsection {\bf Approach II: Factorization models}

Another way to explain the hierarchy of quark and lepton 
masses that has been proposed in the literature \cite{barr1,mohap,barr} and that
does not involve abelian family symmetry, is the idea of
the ``factorization" of fermion mass matrices.
If the Yukawa couplings are matrices in family space,
like $\sum_{ij} F_{ij} (Q_i^c Q_j H)$, then one expects mass
matrices of rank 3 with no particular hierarchy among their
eigenvalues. On the other hand, if one has a factorized
form like $\sum_{ij} F_iF'_j(Q_i^cQ_j H)$, then {\it no matter in what
direction} $F_i$ and $F_j'$ {\it point} the mass matrix
is rank 1. If this were the dominant term it would explain why
one family is much heavier than the others. A second (by
assumption smaller) term $G_i G_j' (Q^c_i Q_j H)$ would,
together with the previous term, automatically give a mass matrix
of rank 2 (unless $G \parallel F$ or $G' \parallel F'$).
And similarly a third term would give mass to the remaining
generation. The crucial point is that no special relative directions
or absolute directions for the vectors 
$F_i$, $G_i$, etc., need be assumed. In a sense, the basic
idea of factorization is opposite to that of abelian continuous or
discrete family symmetry: the whole idea is {\it not} to specify
{\it a priori} special directions in family space. 

The way such factorized forms arise in the models proposed
in the literature \cite{barr1,mohap,barr} is through integrating out
heavy fields. For example, if there are the terms
$M \Psi^c \Psi + F_i \Psi^c Q_i \langle \Omega \rangle 
+ F'_i d^c_i \Psi \langle H \rangle$, where $M$ and
$\langle \Omega \rangle$ are comparable and large, then
integrating out $\Psi^c + \Psi$ gives the effective term
$F_i F'_j (\langle \Omega \rangle/M) Q_i d^c_j H$.

It is easy to see that one may implement this factorization
idea readily in the context of SO(3) family gauge symmetry.
One simply has to promote the Yukawa couplings that are
vectors in family space to be the VEVs of Higgs fields that
are triplets of SO(3). One actually achieves more flexibility
in doing this, because in the models \cite{barr1,mohap,barr} where the
Yukawa vectors are just numbers, they must have (barring
fine-tuning) arbitrary relative directions, whereas here
more than one ``Yukawa" vector could come from the same 
SO(3) Higgs triplet and so be aligned.

Here we will present an SO(10) model with SO(3) gauge family symmetry
that is similar to that proposed in Ref. \cite{barr}, which explains
in an economical way many features of the quark and lepton 
spectrum, and which solves the problem of FCNC coming from
squark-mass non-degeneracy. For helpful background the reader is 
referred to Ref. \cite{barr}.

In this example model, as in Ref. \cite{barr}, the 
nontrivial pattern of quark and lepton masses
is a consequence of the mixing of the three families
with new vectorlike family-antifamily pairs. In particular, we
posit the existence of an SO(3) triplet of families, ${\bf \vec{16}}$,
and a set of fields which are singlets under SO(3) but
transform under SO(10) as the vectorlike pairs 
${\bf 16} + {\bf \overline{16}}$, 
and ${\bf 16'} + {\bf \overline{16}'}$, and the real
representations ${\bf 10} + {\bf 10'}$. 
Involved in the
Yukawa terms of the quarks and leptons are the Higgs fields ${\bf
45}_H$, ${\bf 45}_X$, ${\bf 45}_{B-L}$, ${\bf 16}_H$, and ${\bf 10}_H$.
The two adjoint Higgs point respectively in the $X$ and $B-L$
directions, where $X$ is the SU(5)-singlet generator. 
(The field ${\bf 45}_H$ which is assumed to have a VEV in a general direction 
need not be distinct from ${\bf 45}_X$, but its VEV cannot be strictly
along the $B-L$ direction, see below.)  The
${\bf 16}_H$ gets a GUT-scale VEV in the SU(5)-singlet direction
and a weak-scale VEV in its ordinary weak-doublet component.
The ${\bf 10}_H$ also breaks the weak interactions.

The Yukawa terms of the quarks and leptons are as follows:
\begin{equation}
\begin{array}{lllllll}
{\cal L}_{\rm Yukawa} & = & a_0 M {\bf 16} {\bf \overline{16}} 
& + & (\vec{a} \cdot {\bf \vec{16}})
{\bf \overline{16}} \langle {\bf 45}_H \rangle & + &
(\vec{a} \cdot {\bf \vec{16}}) {\bf 16} \langle {\bf 10}_H \rangle \\
& & b_0 {\bf 16'} {\bf \overline{16}'}
\langle {\bf 45}_X \rangle & + & (\vec{b} \cdot {\bf \vec{16}})
{\bf \overline{16}'} \langle {\bf 45}_{B-L} \rangle & + &
(\vec{b}' \cdot {\bf \vec{16}}) {\bf 16'} \langle {\bf 10}_H \rangle \\
& & c_0 {\bf 10} {\bf 10'} \langle {\bf 45}_X \rangle & + & 
(\vec{c} \cdot {\bf \vec{16}}) {\bf 10} \langle {\bf 16}_H
\rangle & + & (\vec{c^\prime} \cdot {\bf \vec{16}}) {\bf 10}' \langle {\bf 16}_H
\rangle 
\end{array}
\end{equation}
\noindent
The vectors $\vec{a}$, $\vec{b}$, $\vec{b}'$, $\vec{c}$, and
$\vec{c^\prime}$ at this level are just coupling constants which 
explicitly break SO(3), but they are to be understood
as arising from the VEVs of SO(3)-vector Higgs fields in an 
SO(3)-invariant theory.
The same vector $\vec{a}$ appears in two terms, and can be
taken without loss of generality to point in the `3' direction.
(These vectors are real, for reasons explained above, to order
$10^{-4}$.) The two vectors $\vec{b}$ and $\vec{b}'$ are 
assumed to be coplanar with $\vec{a}$ and can without loss of 
generality be
taken to lie in the `2-3 plane'. The vector $\vec{c}$ (and/or
$\vec{c^\prime}$) has a component in the `1' direction.
This pattern will be shown to arise naturally from the underlying
theory with SO(3) family symmetry.

The first set of three terms in Eq. (17) generates mass for the third
generation. Integrating out the ${\bf 16} + {\bf \overline{16}}$
gives an effective term of the form $(Q \vec{a} \cdot {\bf \vec{16}})
(\vec{a} \cdot {\bf \vec{16}}) \langle {\bf 10}_H \rangle/ a_0$,
where $Q$ is the generator of SO(10) that tells which direction
$\langle {\bf 45}_H \rangle$ points in. It is easy to show \cite{barr}
that, no matter what $Q$ is, $m_b = m_{\tau}$ at the GUT scale. 
The second set of three terms generates the next layer of the
onion, namely masses for the second generation. 
Integrating out the ${\bf 16'} + {\bf \overline{16}'}$ one obtains
an effective operator approximately (for small mixing) 
of the form $(\frac{B-L}{X} \vec{b} \cdot
{\bf \vec{16}}) (\vec{b}' \cdot {\bf \vec{16}})$. For fermions
of type $f$ ($f = u,d,l$) this term gives
$(M_f)_{ij} \propto \left( \frac{B-L}{X} \right)_{f^c} b_i b_j'
+ \left( \frac{B-L}{X} \right)_f b_j b_i'$. For the down-quark and
charged-lepton mass matrices, this gives non-zero contributions
to the (22), (23), (32), and (33) elements that are all of the same order
(assuming that the vectors $\vec{b}$ and $\vec{b}'$ point in
arbitrary directions in the (23) plane). The second generation masses,
then, come predominantly from the (22) elements, and $V_{cb} \sim m_s/m_b
\sim \lambda^2$.
Moreover, because of
the factor of $B-L$, the entries in the lepton mass matrix are
three times as large as those in the down-quark mass matrix. This
gives the Georgi-Jarlskog \cite{jarlskog} factor of 3 between the $\mu$ and
$s$ masses.  

On the other hand, because $X$ is the same for $u$ and $u^c$, while
$B-L$ is opposite in sign, 
it is obvious that the contribution of this term to the 
up-quark mass matrix is {\it anti-symmetric}. Therefore the (22)
element of $M_u$ vanishes, and the $c$ mass comes from mixing with the
third generation. Thus $m_c/m_t  \sim \lambda^4 \ll m_s/m_b$, as 
observed. 

Finally, the last set of three terms in Eq. (17) generates the 
first generation
masses and mixings. It is easily shown that the effective operator
gotten from integrating out the ${\bf 10} + {\bf 10'}$ is
(if mixings are small) antisymmetric \cite{barr}. This can be seen from the
fact that, because the ${\bf 45}$ is antisymmetric, the effective
operator changes sign when one interchanges the ${\bf 10}$ and
${\bf 10'}$. This also interchanges $\vec{c}$ and $\vec{c^\prime}$ and
thus the rows and columns of the mass matrices. This antisymmetric  
contribution is only to the down-quark and lepton mass matrices. 
Because the ${\bf 10}$ of SO(10) does not contain up quarks, there is
no contribution from this effective operator to the up-quark mass
matrix.

One has therefore accounted for several more facts. The relative
smallness of $m_u/m_t$ is a consequence of the up-quark matrix still 
being rank-two at this level. The facts that the (11) elements vanish
and that the (12) and (21) elements are equal in magnitude lead both
to the familiar relation $\tan \theta_C
\cong \sqrt{m_d/m_s}$, and to the Georgi-Jarlskog factor of $\frac{1}{3}$
between the $e$ and $d$ masses (since $Det(M_d) \simeq Det (M_l)$).  

Let us now turn to the underlying SO(3)-invariant theory.
There must be at least three SO(3)-vector
Higgs, which again shall be called $\vec{A}$, $\vec{B}$, 
$\vec{C}$, etc.. 
The Yukawa coupling constant $\vec{a}$ comes
from the VEV of $\vec{A}$. So the second term in Eq. (17) comes
from a higher-dimension term $(a/\Lambda) (\vec{A} \cdot
{\bf \vec{16}}) {\bf \overline{16}} {\bf 45}_X$. $a$ is
just some dimensionless coupling, and $\Lambda \sim M_{Pl}$.  
Similarly,
for the third term in Eq. (17). Because the effective Yukawa
coupling of the top quark is about $1$, it is easily seen
that $\langle \vec{A} \rangle \approx \Lambda$. 
The Yukawa constants, $\vec{b}$, $\vec{b}'$, $\vec{c}$,
and $\vec{c^\prime}$ come from the VEVs of $\vec{B}$ and $\vec{C}$.
Because these lead to smaller quark and lepton masses, it
is natural to assume that these VEVs are somewhat smaller
than $\Lambda$. For example, from the earlier discussion
one expects that $\left| \langle \vec{B} \rangle 
\right|^2/\Lambda^2 \sim m_s/m_b \sim \lambda^2$.

Clearly, there must be some symmetry that distinguishes
the three fields $\vec{A}$, $\vec{B}$, and $\vec{C}$,
or else they would couple indiscriminately and   
destroy the specific pattern in Eq. (17). A simple possibility is
$K_A \times K_B \times K_C$, where each $K$ is a $Z_2$
which reflects one of the vectors. By means of this symmetry
one can ensure that at the dimension-four level only
the vector $\vec{B}$ appears in the fifth and sixth term of
Eq. (17). However, at the dimension-six level one can have 
also $(\vec{B} \cdot \vec{A}/ \Lambda^2) \vec{A}$, which is of the
same order as $\vec{B}$, and has the same transformation
properties under $K_A \times K_B \times K_C$. 
Thus the effective Yukawa couplings
in the fifth and six terms of Eq. (17) will be (different) linear 
combinations of $\vec{B}$ and $\vec{A}$. Vectors $\vec{a}$, $\vec{b}$,
and $\vec{b}'$ will be approximately coplanar, and the angles
between them will be order unity, as desired. (There will
also be contributions of order $(\vec{B} \cdot \vec{C}/ \Lambda^2)
\vec{C}$, but these will be negligibly small.) 

In the eighth and ninth terms of Eq. (17) one must have
$\vec{C}$ and another vector $\vec{D}$ (which could be $\vec{B}$) 
at the dimension-four level. Then $\vec{c}$
arises predominantly from $\vec{C}$ and $(\vec{C} \cdot \vec{A}/\Lambda^2)
\vec{A}$, and lies approximately in the $A-C$ plane. 
Similarly, $\vec{c^\prime}$ arises predominantly from $\vec{D}$ and
$(\vec{D} \cdot \vec{A}/\Lambda^2) \vec{A}$ and lies approximately
in the $A-D$ plane. With this arrangement the full rank-three
mass matrices of the quarks and leptons are generated.
From the magnitude of the first generation masses it
is reasonable to assume that $\left| \langle \vec{C} \rangle
\times \langle \vec{D} \rangle \right| \sim \lambda^3$.

It is apparent that the scenario outlined above requires
that the VEVs of the vectors $\vec{A}$, $\vec{B}$, 
$\vec{C}$, etc. must have some non-trivial relative orientation;
they cannot be nearly parallel or nearly orthogonal. 
Superpotentials leading to this situation are easy to construct.
Consider, for example, the angle between $\vec{A}$ and $\vec{B}$.
As noted above, if there are both linear and quadratic 
terms in $(\vec{A} \cdot
\vec{B})$ then a nontrivial angle will in general result. 
However, while the term $(\vec{A} \cdot \vec{B})^2
/\Lambda$ is allowed by $K_A \times K_B$, $M \vec{A} \cdot \vec{B}$
is not. This problem is overcome if there are singlet fields, $A_0$
and $B_0$, that are odd under $K_A$ and $K_B$ respectively, and whose
VEVs are of the same order as $\langle \vec{A} \rangle$ and
$\langle \vec{B} \rangle$. Then a term of the form 
$A_0 B_0(\vec{A} \cdot \vec{B})/ \Lambda$ is allowed.  Another
possibility is to utilize the coupling $\left|\vec{A} \times \vec{B}
\cdot \vec{C}\right|^2$ which is invariant under the discrete
symmetries.  This term will induce non--trivial angles between the
vectors.  Yet another possibility is to use a singlet $\phi$ which is
odd under all
three parity symmetries, this can have an invariant coupling $\vec{A}
\times \vec{B} \cdot \vec{C} \phi$, which will again induce
non--trivial angles between the vectors.

\section{The Lifting of the Squark-mass Degeneracy}

How big will the violations of squark-mass degeneracy be?
We will discuss this in the context of the
SO(10) example presented in the last section.
The dominant contribution to the $\tilde{d}$-$\tilde{s}$
non-degeneracy, in complete analogy with the
``SU(2)" models discussed in Sec. 2, comes from
effective terms of the form
\begin{equation}
{\cal L}_{{\rm soft}} = \int d^4 \theta {\bf \vec{16}}^{\dag}
\cdot \vec{B} \vec{B}^{\dag} \cdot {\bf \vec{16}} \overline{\eta} \eta
/\Lambda^2.
\end{equation}
\noindent
Since we know from the discussion of the last section that
$\left| \langle \vec{B} \rangle \right|^2/\Lambda^2 
\sim m_s/m_b \sim \lambda^2$,
it follows that this term will contribute $O(\lambda^2)$ 
to the (22), (23), (32), and (33) elements of the squark mass matrices,
which is consistent with Eq. (4) and sufficiently small.
This is the same result that obtains in 
the models based on SU(2) and its discrete subgroups.

In the kinds of models described in the last section 
the large VEV of $\vec{A}$ breaks SO(3) down
to an approximate SO(2). Since this is abelian one might wonder whether this 
residual symmetry will be sufficient to protect the
near-degeneracy of the $\tilde{s}$ and $\tilde{d}$ squarks. The dangerous operator
that is allowed by SO(3) (and which has no analogue in
the models based on SU(2) and its subgroups) is
\begin{equation}
\int d^4 \theta ({\bf \vec{16}}^{\dag} \cdot \vec{A} \times {\bf \vec{16}})
\overline{\eta} \eta /\Lambda.
\end{equation}

\noindent
While this is invariant under SO(3), it is odd under $K_A$.
However, at least in the models of Approach II,
an allowed and equally dangerous term is gotten
by multiplying the above operator by $A_0/\Lambda$.

Fortunately, this dangerous class of operators can be forbidden
by going to the gauge group O(3), instead of SO(3). In
other words, by adding the $Z_2$ under which all vectors
of SO(3) are odd. Then the dangerous operator of Eq. (19) can
only be made even under this $Z_2$ by putting in an 
odd number of additional factors of SO(3)-vector fields.
But this, in turn, is only possible using the ``triple product",
$\vec{A} \cdot \vec{B} \times \vec{C}$. Hence, the lowest 
operator of the type shown in Eq. (19) is 
\begin{equation}
\int d^4 \theta ({\bf \vec{16}}^{\dag} \cdot \vec{A}
\times {\bf \vec{16}}) (\vec{A} \cdot \vec{B} \times \vec{C})
B_0 C_0 \overline{\eta} \eta / \Lambda^4,
\end{equation}

\noindent
which is $O(\frac{B^2}{\Lambda^2} \frac{C^2}{\Lambda^2} m_{SUSY}^2)
\sim O(\lambda^5)$
and therefore negligible.

To avoid problems in the Kaon system it is also necessary
to ensure that the mixing of $\tilde{b}$ with $\tilde{s}$
and $\tilde{d}$ be sufficiently small, and in particular
of order $\lambda^2$ and $\lambda^3$ respectively. (See Eq. (5).)
The $\tilde{b}$-$\tilde{s}$ mixing comes both from Eq. (16)
(which we have seen is not dangerous) and from effective operators
of the form
\begin{equation}
\int d^4 \theta {\bf (\vec{16}}^{\dag} \cdot \vec{A}) (\vec{B}
\cdot {\bf \vec{16}}) (\vec{A} \cdot \vec{B} \;\;
{\rm or} \;\; A_0B_0) \overline{\eta} \eta/\Lambda^4.
\end{equation}

\noindent
Note that one had to go to order $\Lambda^{-4}$ to get
an operator that is invariant under $K_A \times K_B$.
This gives a result for the mixing that is of order $\lambda^2$,
as required. (The result is even smaller in Approach I models
where $\vec{A} \cdot \vec{B}$ vanishes (at least approximately)
and where there is no need for the fields $A_0$ and $B_0$.)

The $\tilde{b}$-$\tilde{d}$ mixing comes from
operators of the same form as Eq. (21) with $\vec{B}$ and $B_0$
replaced by $\vec{C}$ and $C_0$. The resulting mixing
is typically of order $\lambda^3$ as required.

\section{The Breaking of CP and Electric Dipole Moments}

So far we have an explanation using spontaneously broken
SO(3) and CP for the apparent near-degeneracy of the 
squarks of the first two generations. Since that explanation
involves certain CP-violating phases being naturally small,
one can get ``for free", as it were, a solution to problems
(2) and (3) discussed in section 2. However, this does not
solve problem (4), since the family SO(3) does not
make the A-terms proportional to the Yukawa terms.
Moreover, even if the phases in the SO(3)-breaking VEVs,
the squark masses, the A parameter(s) and $\mu$ are naturally
very small, the Kobayashi-Maskawa mechanism requires that
some phase in the quark mass matrices be of order unity.
(There is no conflict here with the assumption that SO(3) breaking
VEVs are approximately real, since order one complex phases can arise
from the SO(10) breaking sector.)  
This would lead in general to $d_n$ being about two
orders of magnitude too large.  As noted already, RGE running can
introduce some proportionality, of order 10\% or so between
$A$ and the Yukawa couplings, so the tuning needed is only about
$1/10$.  This may not be a serious difficulty.  Nevertheless, let us
elaborate on some other alternatives to avoid problem (4).

There are several ways around this problem. The most radical
is simply to abandon the Kobayashi-Maskawa mechanism. 
That is, one lets the phases in the quark mass matrices
be small also. This does not create a problem in explaining
$\epsilon_K$, since the gluino-box diagrams that were the
source of problem (2) can do that --- indeed we have had
to assume that certain phases were about $10^{-4}$ to avoid
$\epsilon_K$ being {\it too large}. However, these box diagrams will not
give $\epsilon'/\epsilon$ near the present experimental limit. 
If $\epsilon'$ is indeed near the present limit, one would have to
invoke some new mechanism. One possibility is that spontaneous
CP violation occurring in the Weak scale Higgs sector
could give milliweak-type effects. Since it is not necessary
that these give a significant contribution to $\epsilon_K$
(which is already assumed to be taken care of by gluino loops)
it would seem possible that Higgs effects could generate
$\epsilon'$ and $d_n$ of the right magnitudes.

It is also possible to maintain the Kobayashi-Maskawa explanation of the
Kaon-system CP violation. According to Ref. \cite{rata} one needs,
in order to solve problem (4), that the quark mass matrices 
(and A-terms) have a 
special form. In particular in is helpful if certain off-diagonal
elements vanish \cite{rata}.  But we have already seen that in Approach I
models can be constructed with non-trivial ``textures" for the
fermion mass matrices. 

It is somewhat trickier to solve problem (4) in models which
are based on Approach II while maintaining the KM mechanism.
However, it is possible to do so, as we shall show by 
a slight modification of the SO(10) model presented above.

Suppose that the effective Yukawa couplings $\vec{c}$ and
$\vec{c^\prime}$ in Eq. (17) arise (at dimension-4 level) from
the same SO(3) vector Higgs, $\vec{C}$. Then both $\vec{c}$
and $\vec{c^\prime}$ lie in the $\vec{A}$-$\vec{C}$ plane (approximately).
To the extent that this approximation is good, the last three
terms in Eq. (17) will only generate (13), (31), (23), and (32) elements.
These elements of the down-quark mass matrix can have phases 
that are large, with reasonable values
of squark masses, without leading to an excessive $d_n$ \cite{rata}. Thus
by introducing a large phase into some VEV that appears only in
one of the last three terms in Eq. (17), one can get a large
KM phase, $\delta_K$, without creating a difficulty.
However, a large phase in the (12) and (21) elements of the quark mass matrices
would be fatal \cite{rata}.  But with the new assumption we have made one can show that
the (complex) contribution to these elements from the
last three terms of Eq. (17) is only of order $\lambda^5$.
One still needs to generate a real contribution to the (12) and (21)
that is of order $\lambda^3$ to give first generation fermion
masses and the Cabibbo angle. This can be done by introducing
a new set of terms analogous to the last three in Eq. (17) but
which do not have a large phase. 

\section{Conclusions}

We have suggested in this paper that a gauged SO(3) family symmetry
can cure the FCNC problem arising from the squark mass
non--degeneracy.  What enabled us to gauge the SO(3) symmetry is the
near--vanishing of the SO(3) D--terms, which potentially could have
contributed at an unacceptable level to the squark mass
non--degeneracy.  Since triplet representation of SO(3) is real, any
non--zero D--term contributions to the squark mass splittings 
can only arise proportional to CP--violating
phases in the triplet VEVs.  If CP violation has a spontaneous origin,
which may be desirable from other considerations (EDM of the neutron and
the electron), these phases can be naturally small.  We showed how it
is possible to transcribe certain recently suggested fermion mass schemes 
based on SO(10) symmetry into this SO(3) framework.  The SO(3) family
symmetry can also mimick Abelian symmetries enabling one to construct
texture models of fermion masses.  

\section*{Acknowledgments}

We wish to thank R. Rattazzi for stimulating discussions.

\end{document}